# Suppression of surface roughening during ion bombardment of semiconductors

*John A. Scott,[1,2,†] James Bishop,[1,†] Milos Toth[1,2]\**

1. School of Mathematical and Physical Sciences, University of Technology Sydney, Ultimo, NSW 2007, Australia

2. ARC Centre of Excellence for Transformative Meta-Optical Systems, University of Technology Sydney, Ultimo, NSW 2007, Australia

\*Milos.Toth@uts.edu.au

ABSTRACT. Ion beams are used routinely for processing of semiconductors, particularly sputtering, ion implantation and direct-write fabrication of nanostructures. However, the utility of ion beam techniques is limited by crystal damage and surface roughening. Damage can be reduced or eliminated by performing irradiation at elevated temperatures. However, at these conditions, surface roughening is highly problematic due to thermal mobility of adatoms and surface vacancies. Here we solve this problem using hydrogen gas, which we use to stabilize surface mass flow and suppress roughening during ion bombardment of elemental and compound semiconductors. We achieve smooth surfaces during ion-beam processing, and show that the method can be enhanced by radicalizing $H_2$ gas using a remote plasma source. Our approach is broadly applicable, and expands the utility of ion beam techniques for the processing and fabrication of functional materials and nanostructures.

**Introduction**
Ion irradiation of solids causes crystal damage initiated by ion impacts that produce vacancy-interstitial pairs (Frenkel pairs), and surface roughening due to surface mass flow caused by ion bombardment. Crystal damage can be alleviated by sample heating which enables minimally-intrusive ion beam processing.[1] In particular, heating a semiconductors to above its recrystallisation temperature gives rise to so-called "dynamic annealing" which can prevent ion-induced damage and amorphization of semiconductors through real-time recovery of crystal damage.[2-7] This approach can be superior to annealing after ion irradiation because Frenkel pairs recombine as they are generated, thus preventing the formation of highly stable defect clusters and amorphous layers.[5,8,9] Dynamic annealing is therefore appealing as a means to achieve damage-free ion beam processing of semiconductors. However, whilst compelling, it is used rarely in practice because heat increases the diffusivity of adatoms and surface vacancies, and gives rise to surface mass flow and roughening.[6,7,10-19] The roughening has been studied in detail, but there are currently no general means to eliminate it.[13] Here, we solve this problem by performing ion beam processing in the presence of a chemical precursor gas – namely, hydrogen – which inhibits surface mass flow and suppresses surface roughening.

    We chose hydrogen because it prevents oxide formation and it can immobilize surface species such as adatoms and vacancies. Indeed, hydrogen is highly reactive, amphoteric, and well-known to passivate dangling bonds and form complexes with vacancy defects.[20-24] The mobilities of vacancies and hydrogen-vacancy complexes have been studied extensively.[25-28]

---

[†] John A. Scott and James Bishop contributed equally to this work.

In the case of metals, hydrogen has been shown to increase the mobility of vacancy defects,[28, 29] *whilst* in semiconductors hydrogen retards vacancy diffusion.[22, 30, 31] Here, we exploit this effect as a chemical means to stabilize surface mass flow during ion irradiation. We achieve smooth surfaces during ion beam processing of crystalline elemental and compound semiconductors by using the combination of sample heating and hydrogen. In the absence of heating, crystal damage is observed, and in the absence of hydrogen, surface roughening is observed. We show that the mechanism of mass-flow-stabilization is a rate-limited chemical process which can be enhanced by radicalizing hydrogen gas by a remote plasma source. Our findings improve understanding of mass flow dynamics during ion irradiation, and expand the usefulness and applicability of ion beam techniques to processing and fabrication of functional materials and nanostructures.

**Results and Discussion**
Ion irradiation was performed using a focused ion beam (FIB) microscope equipped with a scanning electron microscope (SEM), a heating stage and a custom-built gas injection system (Figure 1a). The injector was used to deliver $H_2$ to the sample, and the gas was optionally radicalized using a capacitively-coupled plasma source (see the Supporting Information, Section S1 and Figure S1). Gas injectors are used routinely in focused ion/electron beam systems,[32] and plasma injectors have been used previously in electron microscopes.[33-35] In the present work, the $H_2$ gas is used to immobilize adatoms and surface vacancies, so as to stabilize mass flow and suppress surface roughening during ion beam processing, as is illustrated schematically in Figure 1b,c.

We start by demonstrating the ability of hydrogen to suppress surface roughening. Figure 2 shows atomic force microscopy (AFM) images of <001> oriented Ge, <100> GaAs and <100> GaP, after irradiation by $Ar^+$ ions in three environments: (a-c) vacuum, (d-f) $H_2$ gas and (g-i) $H_2$ plasma. The Ge, GaAs and GaP were maintained at 350, 410 and 410 °C, respectively, during ion irradiation (see methods for details) so as to maintain crystallinity during ion irradiation. The 2D root-mean-square surface roughness ($S_q$) measured from each image is shown in the figure. In the case of Ge, $S_q$ decreased from > 1 nm in vacuum to ~2 Å in the presence of $H_2$ gas. Similarly, for GaAs, and GaP, $S_q$ decreased from ~ 5 nm and ~ 2 nm in vacuum, to ~2 Å and ~7 Å in $H_2$, respectively, demonstrating that hydrogen suppressed surface roughening in all cases.

The plasma results (Figure 2g,h,i) are discussed below. First, we note that the surface roughness observed in vacuum (Figure 2a-c) is in the form of highly symmetric nanopatterns. The patterns arise spontaneously, and have been studied previously in a number of semiconductors,[6, 7, 10-13] including Ge and GaAs.[6, 7, 10] Critically, the patterns are evidence of crystallinity – they form due to diffusion of adatoms and surface vacancies on a crystalline semiconductor, modulated by energy barriers associated with topological surface features and the underlying crystal structure.[11, 14-19] The patterns form only if amorphisation of the semiconductors by ions is prevented by dynamic annealing above the recrystallisation temperature (which is typically approximately one third of the melting point). The objective of the present study is to eliminate the patterns, and more specifically to eliminate the associated surface roughness.

Previously, the patterning seen in Figure 2a-c has been described as a form of "inverse epitaxy" driven primarily by the generation, diffusion and pinning of surface vacancies.[6, 7, 11, 16] This is illustrated schematically in Figure 1b for a temperature at which thermal energy is sufficient for the diffusion of vacancies along a surface, but the energy barrier for diffusion over a step edge (the so-called "Ehrlich-Schwoebel" barrier) is prohibitive. Based on this (simplified) model, we expect pattern formation and surface roughening to be suppressed if vacancies and adatoms are immobilized – which is indeed what we observe when ion irradiation is performed in the presence of $H_2$ gas, as is seen in Figure 2d-f for the Ge, GaAs and GaP crystals, respectively.

Next, we discuss each of the semiconductors in Figure 2 in more detail. In the case of <001> Ge, the patterns observed in vacuum (at 350 °C) consist of rectangular prisms with 4 fold symmetry, characterized by the FFT pattern seen in the inset of Figure 2a. The effect of $H_2$ on the patterns is dramatic (Figure 2d). In contrast to the `vacuum' case, the surface is extremely smooth ($S_q$ = 2 Å rather than 1.3 nm) and shows no periodicity. Radicalizing the $H_2$ gas by igniting a plasma inside the gas injector had no additional effect on the surface structure of Ge (Figure 2g) at the ion irradiation conditions used in this experiment. The roughness is approximately the same as when ion bombardment was performed in $H_2$ gas. We attribute this to the reactivity and flow rate of the $H_2$ being sufficient to suppress surface mass flow at the employed ion beam conditions – namely, a relatively low ion beam energy (1 keV) and fluence (1 x $10^{18}$ cm$^{-2}$, see Methods), which we modify and discuss further below.

The pattern on the surface of <100> GaAs (vacuum, 350 °C) consists of nanogrooves (Figure 2b), similar to prior studies performed under these ion exposure conditions.[7, 36] In the presence of $H_2$ gas, the irradiated surface is extremely smooth (Figure 2e, $S_q$ ~ 2 Å rather than ~ 5 nm). However, in contrast to Ge, radicalizing the $H_2$ gas yields an intermediate surface roughness of ~ 1.4 nm. That is, igniting the plasma causes an increase in $S_q$. We attribute this to instability of the GaAs surface under simultaneous exposure by the ions and the plasma, leading to competing processes such as anisotropic chemical etching which can increase $S_q$. Hence, from a practical viewpoint, <100> GaAs under these particular ion beam and gas flow conditions is representative of a material for which $H_2$ gas is superior to a plasma at suppressing surface roughening.

The behavior of <100> GaP is qualitatively similar to that of <001>Ge. Periodic surface patterning is observed in vacuum ($S_q$ ~ 2 nm, Figure 2c), it is suppressed by $H_2$ gas ($S_q$ ~ 7 Å, Figure 2f), and radicalizing the gas has no significant effect on the roughness ($S_q$ ~ 6 Å, Figure 2i). Quantitatively, the precise values of $S_q$ measured on GaP and Ge are different, likely due to variations in the rates of surface mass flow. These rates are material-dependent, they vary with ion beam and plasma parameters, and we explore and discuss them further below.

The experiments in Figure 2 were performed intentionally using ion parameters reported in prior literature on pattern formation in vacuum, above the recrystallization temperature. These conditions (low Ar$^+$ ion energy and flux of a stationary defocused ion beam) are, however, not appropriate for many applications of ion beams. Hence, we next turn to the use of a focused, 12 keV Xe$^+$ beam (7.3 nA) that is rastered over a relatively small surface area of 30 x 15 um – beam parameters representative of those used for focused-ion-beam nanofabrication.

Figures 3a-c show AFM images of <001> Ge after Xe$^+$ ion irradiation was performed at 400 °C in vacuum, $H_2$ gas and plasma environments, respectively. Introducing $H_2$ gas reduced the surface roughness ($S_q$) from 6.6 nm to 4.6 nm, and radicalizing the gas by the plasma reduced it further to ~3 Å. The dramatic nature of the reduction caused by the plasma is seen also in the 1D line profiles shown in Figure 3d. We attributed it to a high chemical reactivity of the plasma,

which prevents surface roughening through rapid, efficient pinning of adatoms and surface vacancies. This implies an increase in the binding energies of surface atoms. The ion beam sputter rate should therefore decrease, as it scales inversely with binding energy.[37] Hence, to confirm that hydrogen does indeed stabilize the surface, we measured the change in sputter rate caused by the gas and the plasma. Figure 3e shows a plot of the depths of pits made in Ge as a result of the ion beam exposures in vacuum, gas and plasma environments. The measured step height is greatest under vacuum conditions and decreases by ~2% and ~38% in gas and plasma environments. This decrease in sputter rate is consistent with the interpretation that hydrogen acts to chemically stabilize the surface by increasing the binding energies of surface atoms.

To confirm the interpretation of chemical surface stabilization, we performed a number of tests designed to eliminate alternate potential mechanisms. First, the vacuum chamber pressure was monitored during ion irradiation. In the 'vacuum' condition, the base pressure was 9 x $10^{-7}$ mbar, it increased to ~7.5 x $10^{-5}$ mbar during $H_2$ gas injection and did not change when the gas was radicalized by igniting the plasma. Hence, the reduction in surface roughness from 4.6 nm to ~ 3 Å and the corresponding reduction in sputter rate (Figure 3e) cannot be caused by a physical mechanism such as a decrease in ion beam energy or current density due to scattering of ions by gas molecules. To confirm this further, we used electron backscatter diffraction (EBSD) to monitor the extent of Ge crystal damage generated by the ions. EBSD images were collected using a 5 keV electron beam at 70° incidence. Figures 4a-c show EBSD images from regions of Ge that were irradiated by the ions at 400 °C in vacuum, $H_2$ gas and plasma, respectively. For comparison, Figure 4d shows an EBSD image from a region irradiated in vacuum at 40 °C. For a material that is crystalline, EBSD images contain Kikuchi band contrast (visible in Figures 4a-c) due to coherent electron scattering from a periodic lattice. The image contrast information is carried by backscattered electrons which have an escape depth distribution with a maximum at 1.4 nm and a tail that approaches zero at ~ 50 nm (see Monte Carlo simulation results shown in Figure S4). Kikuchi band contrast ($C_K$) therefore provides a measure of the crystallinity of the near-surface region that is damaged by ions during our irradiation experiments. The contrast is absent from Figure 4d due to ion-induced amorphisation of the Ge crystal at 40 °C. Conversely, it is present in Figures 4a-c due to dynamic recrystallisation which prevents amorphisation in all three environments – vacuum, $H_2$ gas and plasma. To substantiate this claim further, we quantify $C_K$ using power spectrum analysis. A power spectrum quantifies image contrast as a function of spatial frequency[38, 39], and can therefore be used to measure specific signal and noise components of an image. Figure 4e shows power spectra of the EBSD images in Figures 4a-d. $C_K$ has a strong effect on power at low frequencies, as is indicated by the arrow labeled "Crystallinity" in Figure 4e. It is negligible at 40 °C – i.e., the amplitude of $C_K$ is approximately equal to that of high frequency ($\gtrsim 10^2$ pixel$^{-1}$) noise in the image. Conversely, at 400 °C, $C_K$ dominates the spectra and it is approximately equal in all three environments (vacuum, $H_2$ gas and plasma). Hence, the power spectra show that $H_2$ gas and plasma do not compromise crystallinity at 400 °C. Instead, the gas and plasma suppress surface roughening (Figure 3d) whilst retaining the desired, beneficial "self-healing" effect of dynamic recrystallisation (Figure 4e).

The Ge results presented thus far show that the plasma can either be equally effective (Figure 2a,d,g) or more effective (Figure 3) than $H_2$ gas at suppressing surface roughening. We attribute this variability to differences in surface mass flow rate (R) under various ion irradiation conditions. Specifically, we argue that surface roughening is rate limited by R, and the suppression of $S_q$ is rate limited by the reactivity of the gas molecules. Hence, if R is low relative to the $H_2$ gas reaction rate at the surface, then the gas can suppress surface roughening efficiently and the plasma is expected to have a negligible effect on $S_q$. Conversely, if R is high, then the extra reactivity of the plasma should yield a significant reduction in $S_q$. To test this hypothesis, we measured $S_q$ as a function of ion beam current density in $H_2$ gas and plasma environments (keeping all other parameters fixed). This test is insightful because the current density determines the generation rates and diffusivities of adatoms and surface vacancies, which in turn determine R and $S_q$.

Figure 5 presents the results of this experiment. Figures 5a and 5b show AFM line profiles from regions of Ge that were irradiated in gas and plasma environments, respectively, using a 5 keV $Ar^+$ beam, a fixed ion fluence of 5.4 nC/$\mu m^2$ (see Methods), and current densities (i.e., ion fluxes) of 1.33, 0.62, 0.38 and 0.26 A/$cm^2$. Corresponding AFM images are shown in Figure S3, and plots of $S_q$ versus current density are in Figure 5c. At low current density (0.26 A/$cm^2$), $S_q$ is ~1.5 nm in both cases, and as the current density is increased to 1.33 A/$cm^2$, $S_q$ increases to ~ 5.0 and ~ 1.9 nm in $H_2$ gas and plasma environments, respectively. That is, $S_q$ scales with current density, and the plasma is equally/more effective than the gas at suppressing roughening at low/high current density, respectively. This behavior is as expected, and consistent with our claim that the observed suppression of $S_q$ by hydrogen is a chemical process which inhibits surface mass flow during ion irradiation.

Finally, we discuss the effects of hydrogen on surface composition. A side-effect of our technique is that some hydrogen is expected to remain at and hence modify the sample surface. However, in general, this is not detrimental. Hydrogen processing of semiconductors is used routinely in the passivation of electrically and optically active defects, protection against surface oxidation and as a stabilizing agent during growth.[40, 41] Hydrogenation reactions in semiconductors are typically relatively easy to reverse and preferred over alternatives such as oxide formation (which will take place at some material-specific rate upon exposure to air).[41] In addition, hydrogen is ubiquitous, it is present during transfer and many chemical processing steps of materials, including synthesis; extreme measures are required to mitigate its presence. Hydrogen is therefore the species of choice for chemical surface stabilization without detrimental material modification.

In the case of compound semiconductors, irradiation by energetic ions modifies surface stoichiometry because the mass and binding energy of an atom play a role in displacement and sputtering.[42,43] Hydrogen therefore likely alters the stoichiometry by increasing the binding energies of surface atoms. The net effect may be a beneficial suppression of preferential sputtering since the binding energies of mobile surface atoms (i.e., those associated with the weakest bonds at the surface) are increased via hydrogenation. However, confirmation of any such effects is beyond the scope of the present work. It requires a detailed analysis of surface stoichiometry as a function of parameters such as ion energy, and ion mass relative to that of target atoms comprising a compound semiconductor.[42]

In summary, we developed a chemical process for suppressing surface roughening under the conditions used to mitigate ion beam damage by dynamic annealing. The method is practical and simple – introduction of $H_2$ gas or plasma during ion beam irradiation – and it works over a wide range of ion beam parameters. It is broadly applicable to both elemental and compound semiconductors, and will enhance the utility of ion beam processing techniques.

**Materials and Methods**
**Ion Beam-Induced Surface Patterns (Figure 2)**

A 10 x 10 mm substrate was cleaved from a <001> orientated Ge wafer (*MTI*) and sonicated in acetone and isopropanol for 15 minutes each, and gently purged with $N_2$. It was then loaded onto a custom-built boron nitride restive heating stage housed within a Thermo Fisher Scientific HELIOS G4 dual (ion-electron) beam microscope, and pumped to high vacuum. A *k-type* thermocouple was clipped onto the stage and coated with silver paste to ensure good thermal contact for temperature measurements. The sample was heated to 350 °C, and stabilized. Irradiations were performed with a stationary, defocused, 1 keV $Ar^+$ beam. The aperture in the ion column was selected to provide a 22 nA beam current, and the irradiation time and beam defocus were set to achieve a fluence of ~1 x $10^{18}$ $cm^{-2}$ per exposure site. Following irradiation, the stage was moved to a pristine area of the sample surface and the gas/plasma gun was inserted. Gas injection capillary was previously aligned to the field of view of the ion beam microscope, at a distance of approximately 200 μm above the sample surface. A hydrogen plasma was ignited and stabilized (7.5 x $10^{-5}$ mbar chamber pressure, 60 W power) and an irradiation was performed with the sample surface under constant plasma flow. Following the irradiation, the RF power for the plasma was turned off, the stage was moved to a pristine region and an irradiation was performed under a constant $H_2$ flow. After the irradiation, gas flow was terminated, the plasma gun was retracted and the stage was cooled. A nearly identical procedure was performed using <100> GaAs and <100> GaP (*Stanford Advanced Materials*) samples. The conditions for patterning were 410 °C sample temperature, 1 keV $Ar^+$ 15 nA 1.6 x $10^{18}$ fluence and 410 °C sample temperature, 8 keV, $Ar^+$, 51 nA, 1.2 x $10^{18}$ fluence for GaAs and GaP respectively.

**Surface roughening of Ge (Figures 3,4)**
A 10 x 10 mm Ge substrate was cleaned, loaded onto the heating stage and pumped to base pressure (~9 x $10^{-7}$ mbar). Once at base pressure the stage was heated and stabilized at 400 °C. A square irradiation pattern with dimensions 30 x 15 μm was defined on the substrate surface. The focused ion beam conditions were: $Xe^+$ 12 keV, 7.32 nA. The scan overlap and dwell time were 50% and 1 μs, respectively. The total mill time was 5 minutes. Irradiations were sequentially performed in vacuum, $H_2$ plasma and $H_2$ gas environments. Following the final irradiation, $H_2$ gas flow was stopped and the sample was cooled to room temperature then removed for characterization. A second <100> Ge sample was loaded for the power spectral analysis work (Figure 4). After loading and pumping to base pressure the sample was heated to 400 °C. Identical square pattern irradiations were performed using a focused ion beam in three environments: vacuum, $H_2$ plasma, $H_2$ gas. The shutoff valve for the gas delivery was closed and the gas/plasma injector was retracted. The sample was then cooled to 40 °C to perform a reference irradiation in vacuum. After cooling to room temperature, the sample was removed, loaded onto a 70° pre-tilt holder and pumped to high vacuum. Electron backscatter diffraction characterization was performed on each irradiation.

**Current Density Series (Figure 5)**
Ion beam milling was performed on a <001> orientated Ge sample at 500 °C using a stationary, defocused 5 keV $Ar^+$ beam. The beam current was 17 nA. The irradiation area and time were adjusted to achieve a constant ion fluence (5.4 nC/$μm^2$) for each irradiation but increasing current densities. Different substrates were used for each environment ($H_2$ gas and plasma). Following the irradiation series the sample was cooled in high vacuum then removed for characterization.

**Characterization**
EBSD characterization was performed using a C-Nano detector (Oxford Instruments). The acquisition time was 2000 ms, the electron beam energy and current were 5 keV and 6.4 nA respectively. Power spectrum density analysis was performed on EBSD patterns to ascertain subtle differences in crystallinity. For each EBSD image a radially-averaged fast Fourier transform (FFT) was performed. The azimuthally average, radial intensity was determined and squared for each 1D power spectral density plot. AFM characterization was performed using a Park XE7 AFM. Monte Carlo simulations were performed using the software package CASINO.[44]

**Supporting Information**
Figures showing pfib dual-beam chamber image including gas/plasma injection system with schematic of capacitively-coupled plasma source; Monte Carlo simulation of backscattered electron escape depth; Surface coarsening of Ge versus ion beam current density.

**Acknowledgements**
The authors acknowledge financial support from the Australian Research Council (LP170100150, CE200100010, DP190101058) and Thermo Fisher Scientific. John A. Scott and James Bishop contributed equally to this work.

**Corresponding Authors**
**Milos Toth**: Milos.Toth@uts.edu.au

**Figures and Tables**

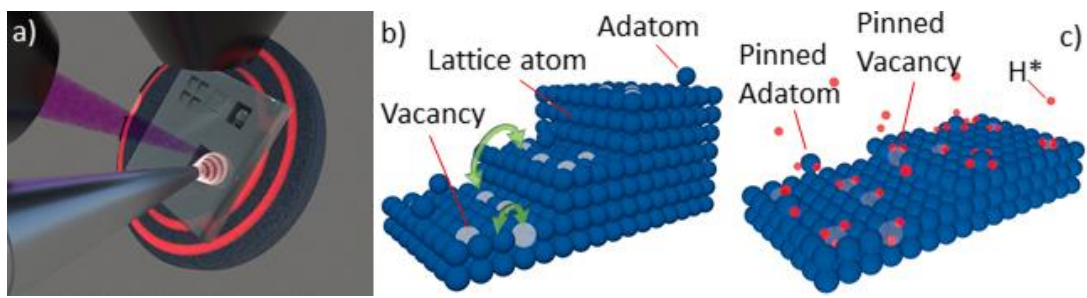

**Figure 1. Ion beam irradiation of a crystalline semiconductor at elevated temperature.**
**(A)** Schematic illustration of a sample mounted on a heating stage of a FIB-SEM instrument equipped with a hydrogen gas/plasma source. The stage is tilted towards the FIB column and irradiated by a hydrogen plasma during ion beam (purple) irradiation. The plasma is shown as a localized plume delivered to the sample by a capillary-style gas injector. **(B)** Diffusion of adatoms and vacancies at the surface of a crystalline semiconductor, resulting in terracing. **(C)** Pinning of adatoms and surface vacancies by hydrogen.

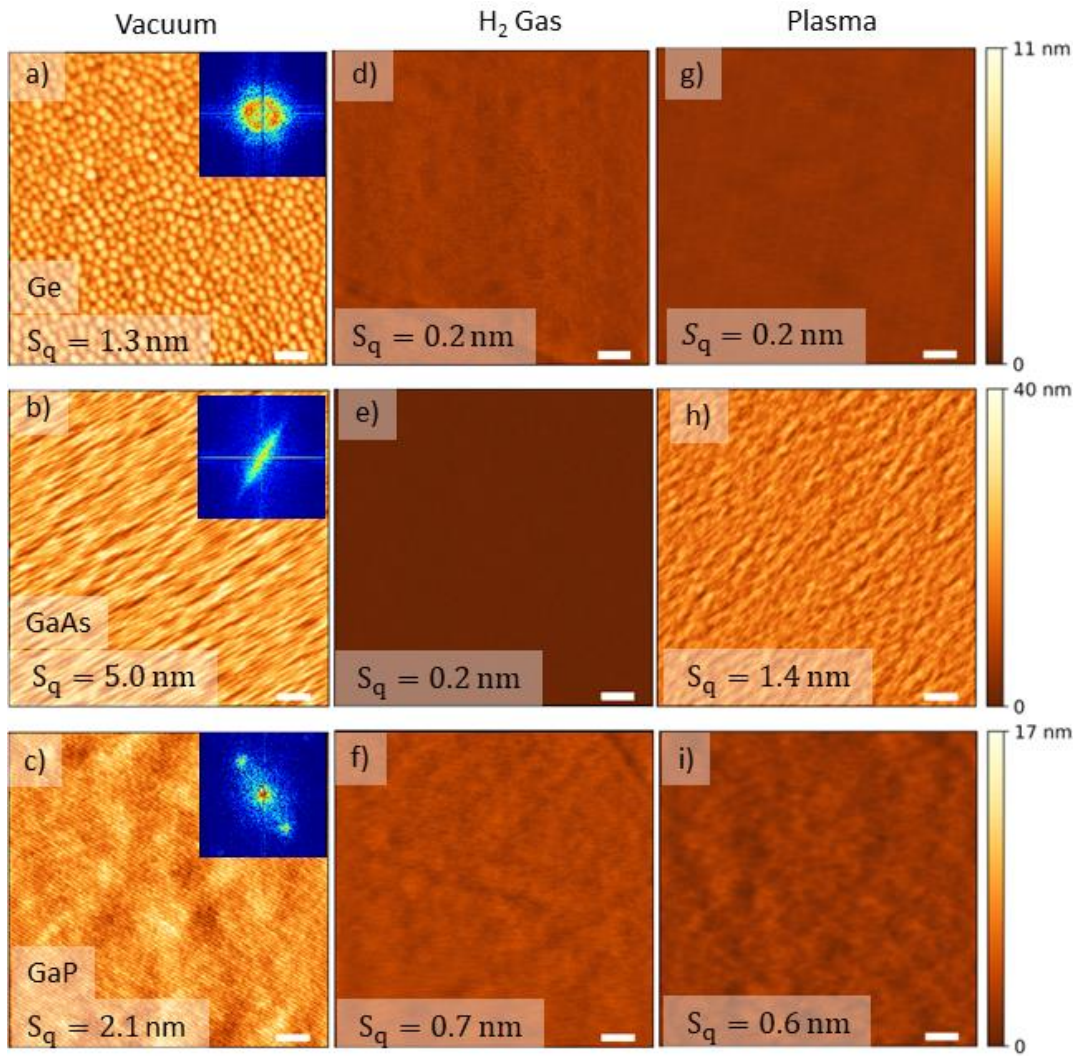

**Figure 2. Surfaces of single crystal semiconductors after ion irradiation at elevated temperature. (A, B, C)** AFM maps of <001> Ge, <100> GaAs and <100> GaP that were irradiated in vacuum, showing symmetric topographic surface patterns that form spontaneously as a result of ion bombardment above the recrystallisation temperature. The insets show FFTs of the AFM maps. **(D, E, F)** Regions of the Ge, GaAs and GaP that were irradiated by the ions in the presence of $H_2$ gas. The surface patterning and roughening that occurs in vacuum is absent in all three cases. **(G, H, I)** Corresponding surface regions that were irradiated by the ions in the presence of a hydrogen plasma (see main text for details). The Ge, GaAs and GaP were maintained at 350 °C, 410 °C and 410 °C, respectively, during ion irradiation. $S_q$ is the 2D root mean square surface roughness. Scale bar = 250 nm.

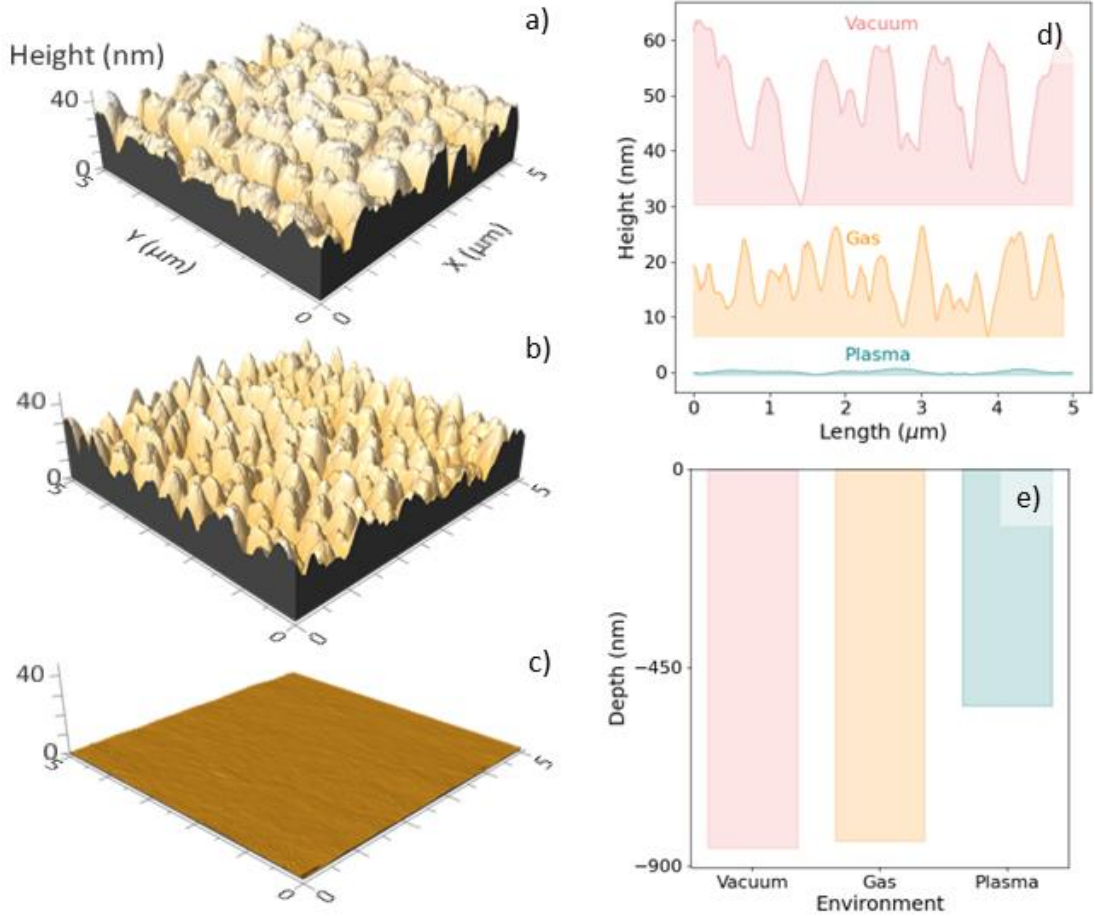

**Figure 3. Surface roughening of Ge due to irradiation by a 12 keV Xe$^+$ ion beam. (A, B, C)** AFM height plots of regions of Ge that were irradiated at 400 °C in: **(A)** vacuum, **(B)** H$_2$ gas, and **(C)** H$_2$ plasma. **(D)** Corresponding line profiles showing the extent of surface roughening in the three environments. The surface roughness ($S_q$) is 6.6, 4.6 and 0.26 nm in vacuum, H$_2$ gas and H$_2$ plasma environments, respectively. **(E)** Plot showing AFM step height measurements of the milled box edge under each environment, showing that hydrogen acts to reduce the sputter rate of Ge. The height measurements are 860 nm for vacuum, 843 nm for gas and 537 nm for plasma.

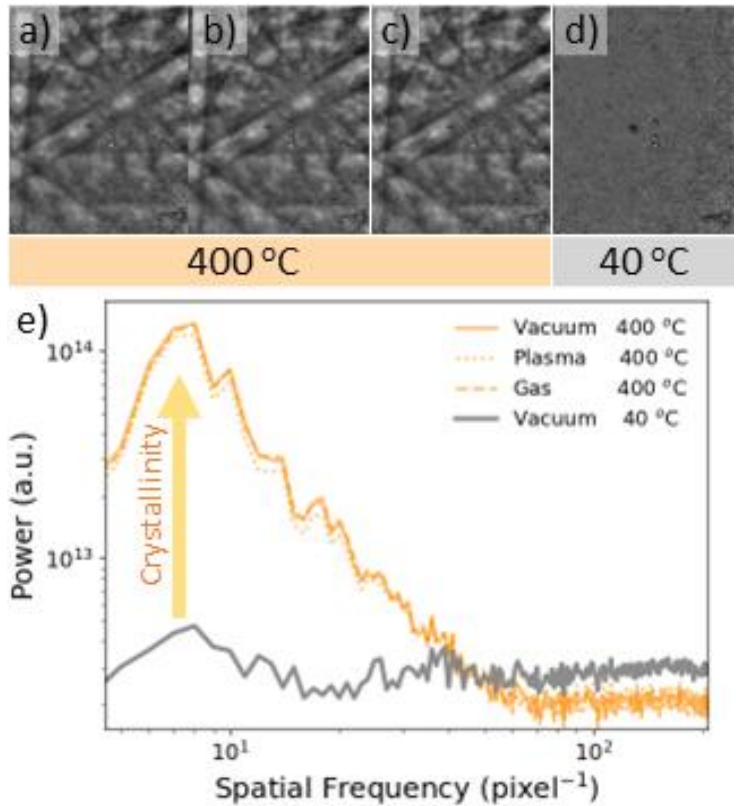

**Figure 4. Crystallinity of Ge irradiated by a 12 keV Xe$^+$ ion beam. (A-D)** EBSD images from regions of Ge that were irradiated in: **(A)** vacuum at 400 °C, **(B)** $H_2$ gas at 400 °C, **(C)** $H_2$ plasma at 400 °C, and **(D)** vacuum at 40 °C. Kikuchi band contrast is absent from image D due to ion-induced amorphisation of the Ge crystal at 40 °C. Conversely, it is present in images A-C due to dynamic recrystallisation which prevents amorphisation at 400 °C. **(E)** Power spectra of the EBSD images. The arrow labeled "Crystallinity" indicates the spectral region that is most affected by Kikuchi band contrast. At 400 °C, the degree of crystallinity is approximately equal in vacuum, $H_2$ gas and plasma environments.

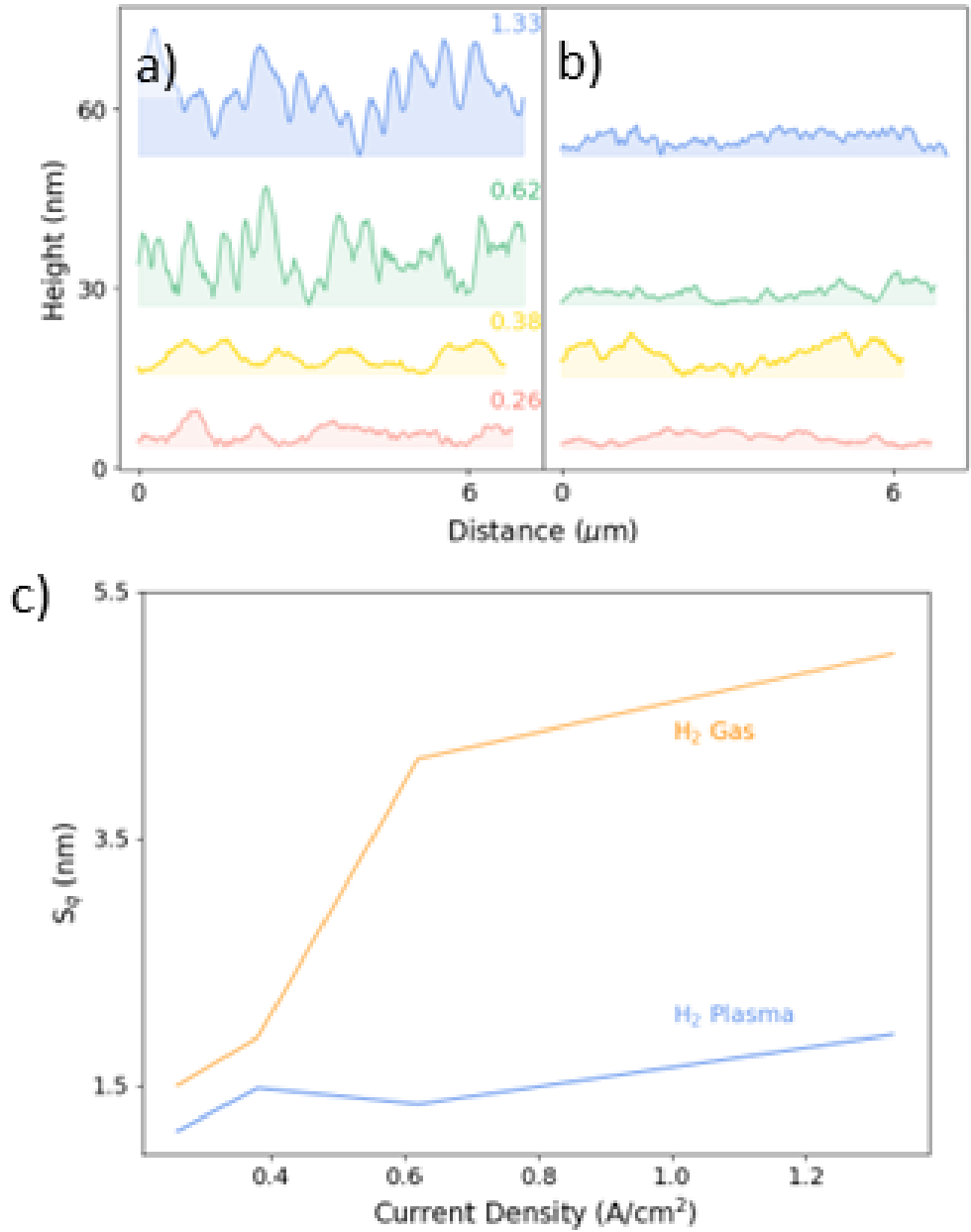

**Figure 5: Surface roughening of Ge versus the current density of a 5 keV Ar$^+$ ion beam in H$_2$ gas and plasma environments. (A,B)** AFM line profiles of surface regions irradiated using a beam current density of 1.33, 0.62, 0.38, 0.26 A/cm$^2$ (top-to-bottom) in: **(A)** H$_2$ gas, and **(B)** plasma environments. **(C)** Corresponding plots of surface roughness (S$_q$) versus current density in H$_2$ gas (orange) and plasma (blue) environments.